\documentclass[useAMS,usenatbib]{mn2e}
\usepackage{scalefnt}
\usepackage[figuresright]{rotating}
\usepackage{subfigure}

\usepackage{natbib}

\title{Imagery and long-slit spectroscopy of the Polar-Ring Galaxy
\mbox{AM\,2020-504}}

\author[Freitas-Lemes, P. et al.]
  {P. Freitas-Lemes$^1$\thanks{E-mail: priscila@univap.br}, I.
Rodrigues$^1$, M. Fa\'undez-Abans$^2$,
 O. L. Dors Jr.$^1$, \newauthor I. F. Fernandes$^3$\\
  $^1$Universidade do Vale do Para\'iba. Av. Shishima Hifumi, 2911, CEP:
12244-000, S\~ao Jos\'e dos Campos, SP, Brazil\\
  $^2$MCTI/Laborat\'orio Nacional de Astrof\'isica, Caixa Postal 21, 
CEP:37.504-364, Itajub\'a, MG, Brazil \\
  $^3$UEFS, Departamento de F\'isica, CEP 44036-900, Feira de Santana, BA,
Brazil}
  
\date{Released 2012 July 26}

\def\cm2{cm$^2$ }
\def\se1{s$^{-1}$ }

\def\arcmin{\hbox{$^\prime$} }
\def\arcsec{\hbox{$^{\prime\prime}$} }

\begin{document}


\maketitle
\begin{abstract}

Interactions between galaxies are very common. There are special kinds of
interactions that produce systems called Polar Ring Galaxies (PRGs), composed by
a lenticular, elliptical, or spiral host galaxy, surrounded by a ring of stars
and gas, orbiting in an approximately polar plane. The present work aims to
study \mbox{AM\,2020-504}, a PRG with an elliptical host galaxy, and a narrow
and well defined ring, probably formed by accretion of material from a donor
galaxy, collected by the host galaxy. Our observational study was based on BVRI
broad band imagery as well as longslit spectroscopy in the wavelenght range
4100--8600\AA,  performed at the 1.6\,m  telescope at the Observat\'orio do Pico
dos Dias (OPD), Brazil. We estimated a redshift of \textit{z}= 0.01683,
corresponding a heliocentric radial velocity of 5045 $\pm$ 23 km/s. The (B-R)
color map shows that the ring is bluer than the host galaxy, indicating that the
ring is a younger structure. Standard diagnostic diagrams were used to classify
the main ionizing source of selected emission-line regions (nucleus, host galaxy
and ring). It turns out that the ring regions are mainly ionized by
massive stars while the nucleus presents AGN characteristics. Using two
empirical methods,  we found  oxygen abundances  for the H\,II regions located
in the ring in the range 12+log(O/H)=8.3-8.8 dex, the presence of an oxygen
gradient across the ring, and that  \mbox{AM\,2020-504}  follows the
metallicity-luminosity relation of spiral galaxies. These results support the
accretion scenario for this object and rules out cold accretion as source for 
the HI gas in the polar ring.
 
\end{abstract}

\begin{keywords}
galaxies: peculiar. galaxies: kinematics and dynamics. galaxies: polar-ring.
galaxies: individual: \mbox{AM\,2020-504}.
\end{keywords}

\section{Introduction}

Galaxies have long been seen as islands distantly scattered and stable in the
universe. We now know that galaxies are not randomly distributed in
space. They are in groups that are subject to the expansion of the universe
and mutual gravitational interaction. The interaction between galaxies has
substantially modified the cosmic structures throughout the evolution of the
Universe. These events are determined by the attractive character of
gravity which in turn induces in larger systems, collisions, tidal forces
and dynamical frictions \citep{1999AJ....117.2695R}. The strong perturbations on
the interacting systems are due to the tidal
force. This can dismember large quantities of material to form bridges
and tails, and thus injecting chemically processed interstellar material into
the 
intergalactic space, contaminating distances up to 10 times larger
than the diameter of the iterating galaxies \citep{1997ASPC..114...71D}.

One of the many types of interactions occur when there
is a ring of gas, dust and stars positioned perpendicularly
with the galaxy's main plane. These systems are known as
polar ring galaxies (PRG), peculiar systems with early-type
or elliptical host galaxies.  The term Polar ring galaxies was first introduced by \cite{1983AJ.....88..909S} and used by Whitmore in the publication \cite{1987ApJ...314..439W}.
\cite{1990AJ....100.1489W} published his ``Polar Ring Catalog" (PRC), with a
total of 157 objects: 6 kinematically confirmed (rotation, detected in two
orthogonal planes), 27 galaxies as ''good candidates", 73 as ``Possible
candidates", and 51 galaxies as ''related objects". \cite{1998ApJ...499..635B} discuss
the origin of the fundamental observational properties of polar ring
galaxies. \cite{1998A&AS..129..357F} make a more
comprehensive classification of all collisional ring galaxies, which
includes the PRGs. Within our neighboring Universe, 20 PRGs have recently
been confirmed in the catalog by \cite{2009Natur.461...43G},  as well as \cite{2011MNRAS.418..244M} display a new catalogue with candidates to polar-ring galaxies selected from the SDSS. 

The \cite{2003A&A...401..817B} reviewed the two scenario for the formation of PRGs: (1) the fusion that occurs in a frontal collision between two spiral galaxies whose discs are orthogonal, (2) the accretion
scenario, in which during the interaction between two galaxies the host collects
material from another galaxy to form the ring. Both scenarios require a
specific geometric configuration for the formation of a polar ring. 

Also, \cite{2006ApJ...636L..25M} proposed the (3) \textit{cold accretion
scenario} for the formation of isolated PRG's. Based on a large cosmological
hydrodynamical simulation, they showed that their formation can occur naturally
in a hierarchical universe where most low-mass galaxies are assembled through
the accretion of cold gas infalling along megaparsecscale filamentary
structures.

Here we report the results of a study of the PRG \mbox{\mbox{AM\,2020-504}},
based
on broad band images and long-slit spectroscopy obtained at
the Observat\'orio Pico dos Dias, Brazil. The main goal of this paper is to
investigate the scenario of formation  by determinations of the oxygen abundance
in the star-forming regions located in the ring and infering  the dust and gas
content of the system. This was done through broadband images and spectroscopic
data, from which we study the kinematics, surface and aperture photometry. In
Section \ref{sec2} we present a review of \mbox{\mbox{AM\,2020-504}}.
Observation and data reductions are presented in Section \ref{sec3}. The results
and discution are presented in Section \ref{sec4}, while the conclusions are
given in Section \ref{con}.

\section{\mbox{AM\,2020-504} review} \label{sec2}

\mbox{AM\,2020-504} is composed by a narrow ring
surrounding a very bright host galaxy. This object appears in many PRG catalogs
\citep[e.g.][]{1990AJ....100.1489W, 2003A&A...401..817B, 2002A&A...383..390R,
1998A&AS..129..357F, 2004A&A...422..941C}. Based on photometric and
spectroscopic observations, \cite{1993A&A...267...21A} concluded that the
material of the ring has been likely accreted, as indicated by the kinematical
decoupling of the inner core of the host galaxy, the different color of the
material in the ring and in the galaxy, and the large amount of HI, quite
unusual for an E galaxy. They modeled the surface brightness of the galaxy,
assuming that the central component is seen edge-on, to determine the geometry
of the system. They found that the luminosity profile of the host galaxy is well
described by an oblate Jaffe model with axis ratio $c/a = 0.6$ for R$>9\arcsec$,
where $c$ and $a$ are the minor and major axis of the galaxy, and $R$ is
the galactic radius.
The intrinsic inclination of the ring plane derived using the \mbox{(B-R)} and
H$\alpha$ images is consistent with the ring being very nearly polar. The ring
is warped and tilted 18$^\circ$ from edge on, passing with the NE side in front
of the elliptical galaxy. \cite{1993A&A...268..103A} reproduced the UV
SED by a model consisting of an elliptical galaxy with a star burst, which give
a lower limit for the age of the polar ring of $1.5\times10^8$ yr, consistent
with the structure being quite young. 

\begin{figure}
  \centering
  \includegraphics[width=\columnwidth]{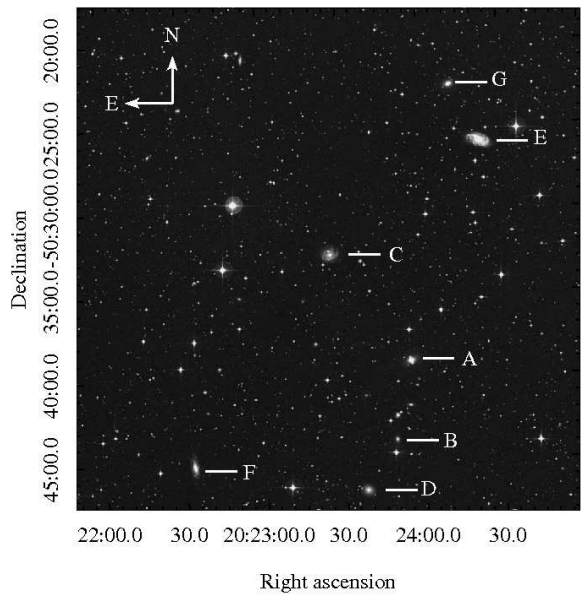}
  \caption{ Objects near the polar ring galaxy \mbox{AM\,2020-504}
(Table\ref{field_vel}). Image obtained from
the DSS.}\label{field} 
\end{figure}

The field near \mbox{\mbox{AM\,2020-504}} is shown in Figure \ref{field}, where
it is
labeled A. \mbox{AM\,2020-504} has no detected leftover
materials or bridges connecting it with another structure or surounding galaxy.
We found six other objects belonging 
to the group in a search field of 30\arcmin around \mbox{AM\,2020-504}.
These galaxies cover a velocity range of 730\,km/s. Closer 
to \mbox{AM\,2020-504}, at a projected distance of $5\arcmin$, is
2MASXJ2023488-5043492
(label B), whose velocity difference is 330\,km/s.  Objects C (ESO\,234-G016)
and D (ESO\,234-G017) have very similar radial velocities (\mbox{190 km/s} and
\mbox{5 km/s}, respectively),
forming a sort of plane in the radial velocity space (see Table\ref{field_vel}).
The coordinates of the objects, as their radial velocities were obtained 
from NED\footnote[1]{NASA/IPAC Extragalactic Database (NED) is operated by the
Jet Propulsion Laboratory, California Institute of Technology, under contract
with the National Aeronautics and Space Administration}.

\begin{table}
\scalefont{0.8}
  \begin{center}
    \caption {Galaxies within a 30\arcmin near \mbox{AM\,2020-504}. Labels are
according to Figure \ref{field}.}
     \label{field_vel}
    \begin{tabular}{|c|l|c|c|c}
       \hline
      \textbf{Name} & \textbf{Label} & \textbf{Velocity} & \textbf{Relative} &
\textbf{Distance}  \\
      \textbf{} & \textbf{} & \textbf{(km/s)} & \textbf{Velocity} &
\textbf{(arcmin)}  \\
      \hline
      \hline

     A &  \mbox{AM\,2020-504}           & 5006$\pm$43 & 0    &  0 \\
     B &  2MASXJ2023488-5043492  & 4676$\pm$45 & -330 & 4.7 \\
     C & ESO\,234-G016          &  5196$\pm$27 & 190  & 7.9 \\
     D & ESO\,234-G017          &  5011        & 5    & 8.3 \\
     E & NGC\,6899              &  5731$\pm$10 & 724  & 13.8 \\
     F & ESO\,234-G013	     &  4786	       & -220 & 14.3      \\
     G & 2MASXJ20241155-5022394 &  5648	       & 642  & 16.7     \\
	   
      \hline
    \end{tabular}
  \end{center}
\end{table}

\section{Observations and data reduction} \label{sec3}

\subsection{Broadband optical imagery}

Photometric observations were performed with the 1.6-m telescope at
Observat\'orio Pico dos Dias (OPD), Brazil 
on July 2008. The telescope was equipped with direct imaging Camera 1, with the
CCD\,106, a back-illuminated 
1024\,x\,1024 detector. 

The data were acquired with standard Johnson B, V, R and I filters. Calibration
was accomplished using repeated observations of standard stars from 
\cite{1992AJ....104..340L} selected fields Mark-A and PG13223-086. The log of
observations is given in Table \ref{obs}. 

Data reductions were performed in the standard manner using the 
IRAF\footnote[1]{Image Reduction and Analysis Facility is developed and
maintained by 
the National Optical Astronomy Observatories (NOAO)} package. This included dark
and 
bias subtraction, and flat-field correction (we have used a mean of several dome
flats 
taken in the appropriate filter). The cosmic rays were removed manually by
masking them 
with the median of adjacent pixels.

\begin{table}
  \begin{center}
    \caption {Log of CCD image observations.}
    \label{obs}
    \begin{tabular}{|c|c|c|}
      \hline
      \textbf{Date} & \textbf{Band-pass} & \textbf{Exp. (s)}\\
      \hline
      \hline

      2008 Jul 07-08 & B & 6 x 300 \\
      2008 Jul 07-08 & V & 6 x 300 \\
      2008 Jul 07-08 & R & 6 x 300 \\
      2008 Jul 07-08 & I & 6 x 240 \\

      \hline
    \end{tabular}
  \end{center}
\end{table}

\subsection{Spectroscopic observations}

The spectroscopic observations were performed with the 1.6-m telescope at OPD
equipped with a Cassegrain spectrograph and CCD105, a back-illuminated
2048\,x\,2048 detector on June 2008 and September 2008. Difraction gratings of
300
lines/mm and 600 lines/mm were used. The aim of the 300 lines/mm grating was to
have a larger spectral coverage (4100--8600\AA), while the 600 lines/mm was
used to obtain higher resolution in the main lines H$\beta$,
[O\,III]$\lambda$5007,
H$\alpha$, [N\,II]$\lambda$6584 and [S\,II]$\lambda$6716,$\lambda$6731.

\begin{figure}
  \centering
    \includegraphics[width=\columnwidth]{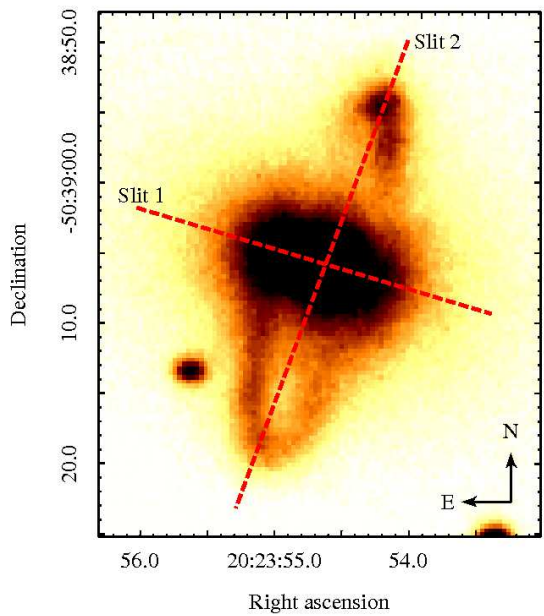}
  \caption{B-band image of the galaxy \mbox{AM\,2020-504} showing the slit
positions.}
  \label{gala}
\end{figure}

For the observations we used two slit positions, one along 
the major axis of the host galaxy (slit-1) and another along the major axis of
the ring (slit-2).
Slit-1 has an inclination of $72.5^\circ$ 
NE and slit-2 has an inclination of $17^\circ$ N-W, as shown in Figure
\ref{gala}. Spectrophotometric standard stars were observed each night to perform flux
calibration. Those are tertiary standards from \cite{1981PASP...93....5B}, as revised by 
\cite{1992PASP..104..533H}, see also 
\cite{1994PASP..106..566H}. Arc lamps were taken before and after each exposure in
order to provide accurate wavelength calibration. A log of the spectral
observations is given in Table \ref{espec}. The spectra processing and data
analysis were done using standard procedures employing IRAF and RVSAO
packages. This includes bias substraction, flat field correction, cosmic
ray removal, sky substraction, wavelength and flux calibration ({\tt
IMAGES}/{\tt IMFIT}, {\tt IMUTIL}, {\tt STSDAS}/{\tt IMGTOOLS}, {\tt TWODSPEC}
and {\tt ONEDSPEC} tasks, respectively). The wavelength calibration
errors were $\simeq 8$\AA\ and $\simeq 10$\AA\ for slits 1 and 2, respectively.
The standard extraction aperture was set for the
emission region. The spectra were reduced using the measurements of the
standard stars observed at
similar airmasses. The line fluxes were obtained using the IRAF/{\tt
SPLOT} task. This task was also used to obtain the center of the emission
lines in order to later calculate the radial velocities of the measured
lines. As a double check of these results, the RVSAO/IRAF external package
was used to calculate the apparent radial velocities from the observed
spectral shifts. The EMSAO task finds emission lines in a spectrum and
computes the observed centers, yielding individual shifts and errors for
each line as well as a single velocity by combining all of the lines
\citep{1995ASPC...77..496M}.

\begin{table}
  \centering
  \caption {Log of spectral observations, the slit positions and exposure
  times.}
  \label{espec}
    \begin{tabular}{|l|c|c}
    \hline
    \textbf{} & \textbf{Slit 1} & \textbf{Slit 2} \\
    \hline
    \hline
    \textbf{Date}   & 2008-Sep-29 & 2008-Jul-04\\
    \textbf{Grating} (lines/ mm)  & 600 & 300\\
    \textbf{Spectral Range} (\AA) & 4600-6730 & 4100-8600 \\
    \textbf{Angle}  & $72.5^\circ$ N-E & $17^\circ$ N-W \\
    \textbf{Exposure Time} (s)  & 1800 & 1200\\
    \textbf{Slit Width} (\arcsec) & 3 & 3 \\
\hline

  \end{tabular}
\end{table}

\section{Results and Discussion} \label{sec4}

\subsection{Rotation Curves}

The radial velocities along the slits were calculated based on the Doppler shift
of spectral lines. To construct the rotation curve of this system, the strongest
emission lines were used, namely:  H$\beta$, [O\,III]$\lambda$5007 and
H$\alpha$. In both slit positions, extractions of one-dimensional spectra were
performed in order to obtain the rotation curves, as well as information on
specific regions.

The radial velocity of the galaxy, calculated by \mbox{averaging} the central
positions of the Slit 1, is \mbox{$5045\pm23$ km/s}. The value is similar to
that found by \cite{1987IAUS..127..413W} and \cite{1993A&A...267...21A}. 

\begin{figure}
  \centering
  \includegraphics[width=\columnwidth]{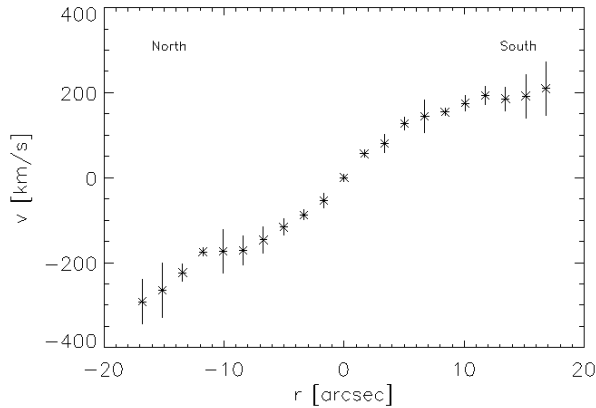}
  \caption{Rotation profile of \mbox{AM\,2020-504} along the ring major axis
(PA=17$^\circ$).}
  \label{rota}
\end{figure}

The rotation profile along the ring major axis is shown in Figure
\ref{rota}. The northern 
portion of the ring is approaching and the southern portion is receding from us.
This rotation curve is symmetrical and well behaved. The last three points each
side of the rotation curve suggest that the northern and southern portions of
the ring have a difference in rotation velocity of about 60 km/s, but this
difference is under the error bars.
To a certain degree, asymmetries could be explained if the ring was warped. In
fact, \cite{1993A&A...267...21A} suggested that the ring is warped, and
generated models that adjust fairly well the morphology and the rotation
velocity curves in some directions for which they had long slit spectra, showing
those asymmetries.

\subsection{Spectral analysis} \label{Section2.2}

To analyse the emission from the gaseous component, we constructed four
diagnostic diagrams proposed by \cite{1981PASP...93....5B} and
\cite{1999A&A...345..733C}, using Slit 2 spectra. These diagrams are used to
distinguish objects ionized only by massive stars from those containing active
nuclei (AGN) and/or shock excited gas.
  The used diagrams were [O\,III]${\lambda5007}$/H$\beta$
versus [O\,I]${\lambda6300}$/H$\alpha$, [N\,II]${\lambda6584}$/H$\alpha$, and 
[S\,II]$({\lambda6717+\lambda6731})$/H$\alpha$; and  
[N\,II]${\lambda6584}$/H$\alpha$ versus 
[S\,II]$({\lambda6717+\lambda6731})$/H$\alpha$. 
The emission-line intensities  were measured using the IRAF SPLOT routine
considering a Gaussian line profile fitting. In Table~\ref{tablen} are
presented the distance to the galactic center (assuming 344 pc/arcsec),
the emission line intensity measurements normalized to the flux of H$\beta$=100,
the nebular reddening coefficient
C(H$\beta$), calculated by comparing the
Balmer decrement H$\alpha$/H$\beta$ to the theoretical value 2.86 given
by \cite{1989agna.book.....O} for an electron temperature of 10\,000 K
and considering the interstellar law of \cite{1958AJ.....63..201W}, and the
observed H$\beta$ flux for each aperture.

The error associated with the line fluxes were estimated following the
same 
procedure than \cite{2008A&A...492..463O}, is given by 
$\sigma^{2}=\sigma^{2}_{cont}+\sigma^{2}_{line}$, 
where  $\sigma^{2}_{cont}$  is the error due to the continuum noise, calculated
to be $\sigma^{2}_{cont}=\sqrt{N}\Delta\sigma_{\rm rms}$,
$N$ is the number of pixels covered
by the emission line, $\Delta$ is the dispersion of the spectrum (units
of wavelength per pixel), and $\sigma_{\rm rms}$ is the root mean square of
the continuum flux density (flux per unit wavelength); and $\sigma^{2}_{line}$
is the Poisson error of the emission line.
 
We have compared the values of the  reddening coefficient C(H$\beta$) in
the ring of \mbox{AM\,2020-504}
with the ones in other PRGs and in isolated galaxies. We found an averaged
C(H$\beta$) of 0.8, which 
is similar to that ($\approx0.9$)  in the ring galaxy 
SDSSJ075234.33+292049.8 \citep{2010MNRAS.401.2067B}, and larger than the ones in  disks
of  
spiral galaxies, for sample, in M\,33 is $\approx0.4$  \citep{2011ApJ...729...56B}
and in M\,101 is $\approx0.4$
\citep{1996ApJ...456..504K}.

\begin{figure}
  \centering
  \includegraphics[width=7cm,angle=270]{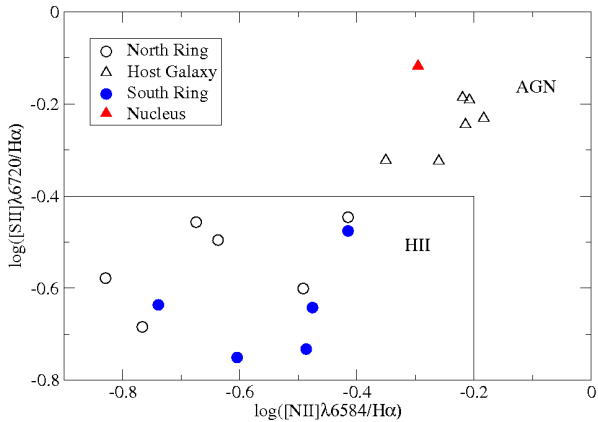}
\caption{\mbox{AM\,2020-504} diagnostic diagram
log[NII]${\lambda6584}$/H$\alpha$ versus
log[SII]$({\lambda6717+\lambda6731})$/H$\alpha$ (Coziol, 1999).  The black
triangle (nuclear region) and 
the white triangle both correspond to the host galaxy. The black and white
circles correspond to the southern and northern regions of the ring,
respectively.}
  \label{diag_cozi} 
\end{figure}

\begin{figure}
  \centering
  \includegraphics[width=\columnwidth,angle=270]{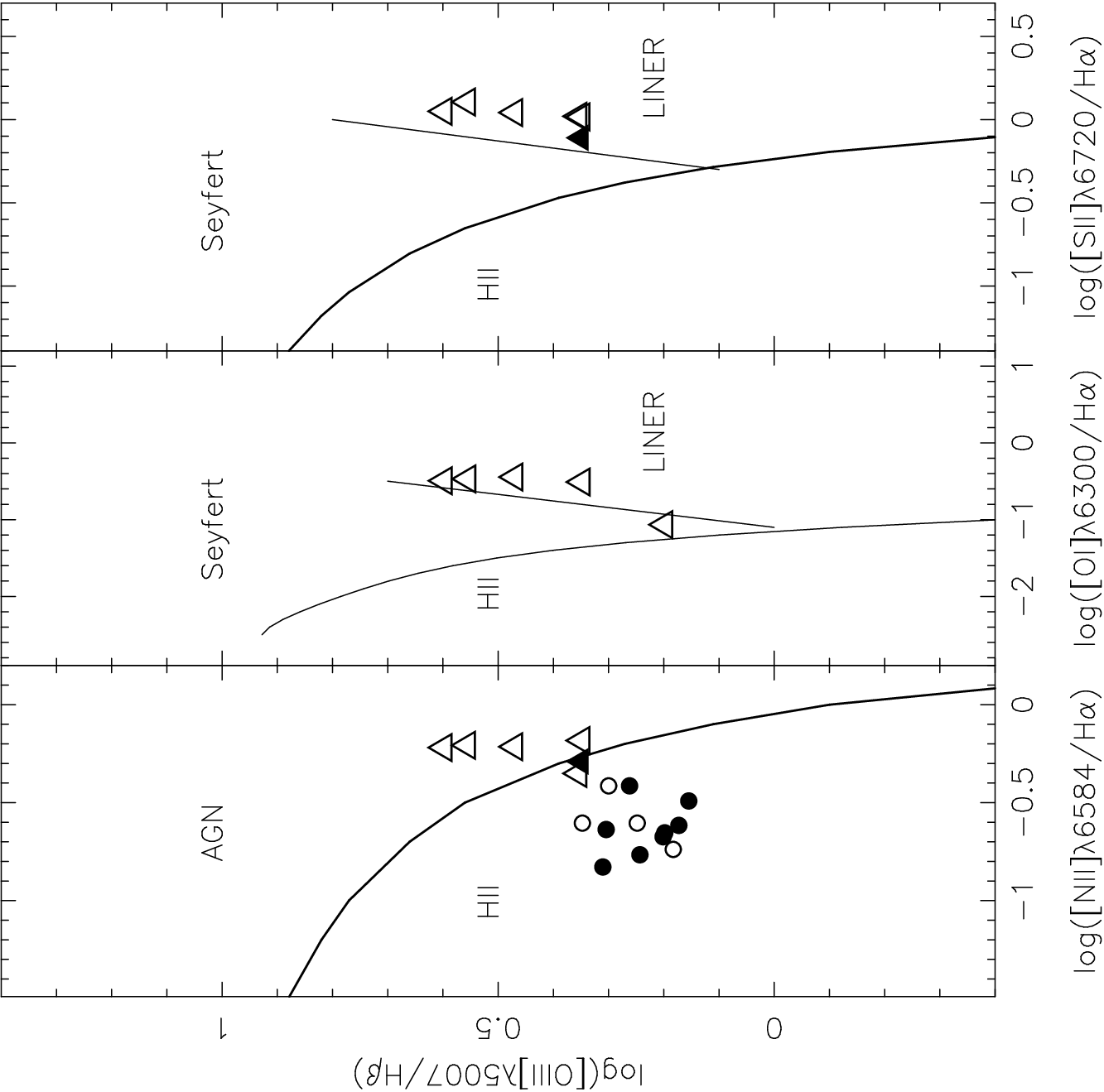}
\caption{Diagnostic diagrams [N\,II]${\lambda6584}$/H$\alpha$ (left),
[O\,I]${\lambda6300}$/H$\alpha$ (middle), and 
[S\,II]$({\lambda6717+\lambda6731})$/H$\alpha$ (right) vs.
[O\,III]${\lambda5007}$/H$\beta$ (Baldwin et al. 1981). The curves, taken from
Kewley et al. (2006) separate objects ionized by massive stars from the ones
containing active nuclei and/or shock excited gas. The straight line taken from
Kewley et al. (2006), separates Seyfert and LINER objects. The symbols are as in
Figure \ref{diag_cozi}}
\label{diag_oli} 
\end{figure}

In Figures \ref{diag_cozi} and \ref{diag_oli} 
the diagrams are shown, where different symbols are used
to represent the ring and the nuclear regions.
Figure \ref{diag_cozi} shows the diagnostics diagram proposed by
\cite{1999A&A...345..733C}, where we plot the values of the [NII]/H$\alpha$
against [SII]/H$\alpha$ ratios. These two line ratios are significantly higher
in LINERs and Seyfert2s than in H\,II regions and Starbursts. The distinction
between the two AGN types (e.g. Seyferts and LINERs) is not possible  in this
diagram, but regions undergoing photoionization by O and B stars are clearly
separated from AGN ionizing sources. The \cite{1999A&A...345..733C} criteria
established two regions in this diagram, separated by the continuous lines in
Figure \ref{diag_cozi}, where the gas is excited by the two different
mechanisms, e.g. AGN and photoionization by stars, which is consistent with the
lower limits for the presence of diffuse ionized gas in the halos of edge-on
Starbursts as proposed by \cite{1996ApJ...462..651L}. The points corresponding
to the ring (circles) are broken down between the northern part of the ring
(open) and South of the ring (closed), the host galaxy is represented by open
triangles and the nucleus by closed triangle.  
 
The lines in the diagrams of Figure \ref{diag_oli} were used to separate objects
with distinct ionization sources, following \cite{2006MNRAS.372..961K} criteria.
These authors combined both photoionization model results  with stellar
population synthesis models, built by \cite{2001ApJ...556..121K}, in order to
analyse the host properties of 
about 80\,000 emission-line galaxies selected from the Sloan Digital Sky Survey.
They showed that Seyferts and low-ionization narrow emission-line regions
(LINERs) form clearly separated branches on the standard optical diagnostic
diagrams, such as the ones used in this paper. We can see that the nuclear
points occupy the AGN's site in four diagrams (see Figures \ref{diag_cozi} and
\ref{diag_oli}), in two of them, these points are at the LINERs site, while the
ring points are in the region occupied for H\,II-like objects.

A fundamental subjects in galaxy formation studies is  our understanding
on the metallicity. In particular,   chemical abundances of H\,II regions in
polar ring galaxies
 have  deep implications for the evolutionary scenario of these objects and
yield hints on the mechanisms
 at work during their formation.
 Three main formation processes have been proposed (for a more detailed
discussion see \citealt{2011A&A...531A..21S} and reference therein):
\begin{itemize}
\item Cold accretion of pristine  gas ---  In this scenario,   a polar structure
may be formed by cold gas accretion 
and it is expected a gas phase with very low  metallicity $Z\sim1/10 Z_{\odot}$.
The metallicity of the galaxy
would then be lower than that of spiral disks of the same luminosity, and any
metallicity gradient along the polar ring would not be present
(\citealt{2006ApJ...636L..25M}, \citealt{2009MNRAS.397L..64A}).
\item Major dissipative  merger --- The PRG is formed from a merger of two disk
galaxies unequal mass (e.g. \citealt{1997ApJ...483..608B}).
\item Tidal  accretion of material --- The polar ring  may be formed by the
disruption of a dwarf companion galaxy or tidal accretion of gas stripping
from a disk galaxy. 
\end{itemize}

In both, major merger and tidal accretion, a somewhat high metallicity would be
found.
To test which of these scenarios represent the formation of \mbox{AM\,2020-504},
  the oxygen abundance
have been estimated in the polar disk regions.
 Unfortunately, accurate chemical
abundances can only be derived by measuring temperature sensitive
line ratios, such as  [O\,III]$(\lambda4959+\lambda5007)/\lambda4363)$,
which are   unobservable  in the spectra of the  H\,II regions in the ring
of \mbox{AM\,2020-504}. In these cases, empirical calibrations between
abundances
and more easily measured emission-line ratios have to be used
to estimate metal abundances (see \citealt{2011MNRAS.415.3616D} and reference
therein). Therefore, we
estimate the oxygen abundance O/H (used as tracer of the metallicity)  using the
calibration of O/H with the  parameters
proposed by \cite{2009MNRAS.398..949P} 

\begin{equation}
O3N2= \rm \log\left(\frac{I([O\:III]\lambda5007)}{I(H\beta)} \times
\frac{I(H\alpha)}{I([N\:II]\lambda6584)}\right),
\end{equation}
\begin{equation}
N2=\rm \log\left(\frac{I([N\:II]\lambda6584)}{I(H\alpha)}\right)
\end{equation}
and given by 

\begin{equation}
{\rm 12 + \log(O/H)} = 8.73 - 0.32 \times O3N2, 
\end{equation}

\begin{equation}
{\rm 12 + \log(O/H)} = 0.57\times N2+9.07.
\end{equation}

\begin{figure}
\centering
\includegraphics[width=\columnwidth,angle=270]{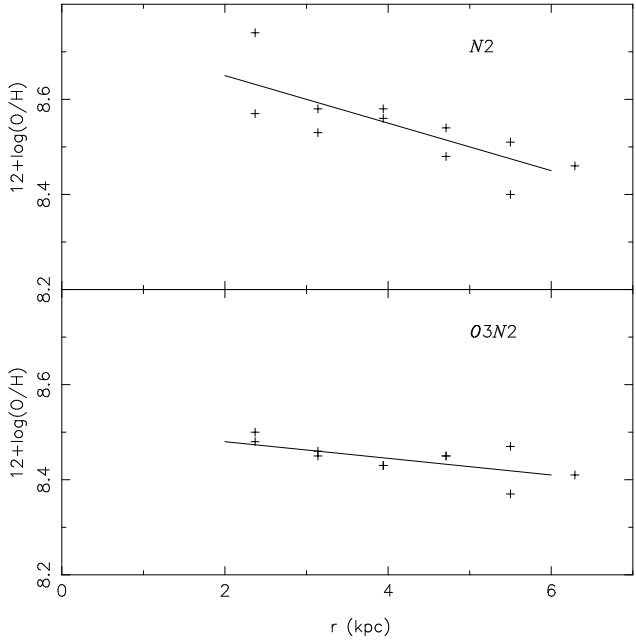}
\caption{Gradients of 12+log (O/H)  in \mbox{AM\,2020-504}. The solid
lines are linear regressions  on oxygen abundance determinations obtained by the
 the parameters indicate in each plot.}
\label{gradi}
\end{figure}

In Table~\ref{tablen} the values of these parameters and the derived oxygen
abundance for each region classified in the
diagnostic diagrams as H\,II regions are presented. We found that $O3N2$ 
and $N2$ parameters indicate  the ring H\,II regions have oxygen abundances 
[12+log(O/H)]
from 8.3 to 8.8 dex, with an  average oxygen value of $8.53\pm0.11$ dex. This
value is about the the solar oxygen abundance, i.e. 8.66 dex 
(\citealt{2004A&A...417..751A}), and near to  the maximum oxygen abundance value
derived for central parts of spiral galaxies, i.e. 8.87 dex
(\citealt{MNR:MNR11444}). 
In Figure~\ref{gradi} the oxygen abundance via the two parameters presented
above as a function of the  galactocentric distance of \mbox{AM\,2020-504} is
shown. We can see that both parameters indicate an  oxygen  gradient across the
ring.
A linear regression in the oxygen estimates via $O3N2$ and $N2$ yields gradients
of   \mbox{$-0.017(\pm0.006$) dex/kpc} and 
 \mbox{$-0.051(\pm0.013$) dex/kpc}, respectively. These values are similar to
the ones found in spiral galaxies (see \citealt{2005A&A...437..837D},
\citealt{2004A&A...425..849P}).

We also tested whether \mbox{AM\,2020-504} follows the 
metallicity-luminosity relation of spiral galaxies.  \cite{2004A&A...425..849P}
found that 
  characteristic oxygen abundance in spiral galaxies
as a function of absolute blue magnitude $M_{\rm B}$ follows the relation
\begin{equation}
12 + \log({\rm O/H}) = 6.93 (\pm 0.37)- 0.079 (\pm 0.018)M_{\rm B}.
\end{equation}
We computed the absolute magnitude of \mbox{AM\,2020-504}, evaluated
considering the central spheroid and the polar ring for a distance of 71 Mpc
(H$_0$=71, \citealt{2000ApJ...529..786M}), being
 $M_{\rm B}$=-18.24. Using this value, we obtained from the relation above 
12+log(O/H)=8.37$\pm0.2$ dex, about the same
 average oxygen value found in the polar ring.

Since the averaged metallicity in  \mbox{AM\,2020-504} is high ($Z\approx
Z_{\odot}$), 
there is a clear  oxygen gradient across the polar ring, and this galaxy 
follows the metallicity-luminosity relation of normal spiral galaxies, our
results 
support the  formation scenarios of  accretion or major merger  for this object
and rule out  the  cold accretion of pristine  gas.

Some other works have determined the oxygen abundance in PRGs in order to
test possible formation scenarios for these
objects. For example, \cite{2002AstL...28..443S}, using a O/H-$N2$ calibration
found the oxygen abundance of 12+log(O/H)$\sim8.8$ dex
for the PRG UGC\,5600.  \cite{2010MNRAS.401.2067B}, using a calibration between
the electron temperature and strong oxygen emission-lines, 
found  that oxygen abundance  in different regions of the apparent ring galaxy
SDSSJ075234.33+292049.8
  is 12+log(O/H)=8.49$\pm±$0.08 dex. \cite{2010ApJ...714.1081S} derive the
oxygen abundance  in the polar
disk of NGC4650A  by using both the empirical  methods and direct electron
temperature detections and found  12+log(O/H)=$8.2\pm± 0.1$.
Recently \cite{2011A&A...531A..21S}, using the $P$-method
\citep{2001A&A...369..594P},    
reported  averaged   oxygen abundance values of $8.5\pm0.5$ and $7.7\pm1.0$
for H\,II regions located in the ring of the galaxies PRG UGC\,7576 and
UGC\,9796.
 Despite of these results are in consonance with our estimations,
 for the majority of the galaxies above the absence of any metallicity gradient
along the polar disk have been found.
 Since different methods or different calibrations of the same oxygen
indicator provide
different oxygen values, with discrepancies of up to 1.0 dex (e.g.
\citealt{2008ApJ...681.1183K}),
and few PRGs have been observed,  additional analysis is needed to
confirm the (dis)aggrement found above.

\begin{table*}
\scalefont{0.8}
\centering
\centering
\caption{Reddening corrected emission line intensities (relative to
H$\beta$=100) and global  properties}
\begin{tabular}{llllllllllll}
 \hline 
\noalign{\smallskip}
\hline
\noalign{\smallskip}
 r    & log($F$(H$\beta$))        & C(H$\beta$)                  &
[O\,III]$\lambda$5007 &   H$\alpha$   &    [N\,II]$\lambda$6584 &
[S\,II]$\lambda$6716 & [S\,II]$\lambda$6731 &  O3N2 &
 N2  & 12+log(O/H)     &    12+log(O/H)   \\
    (kpc)             &          $\rm erg\:s^{-1}\:cm^{-2}$       &             
                    &                     &                                    &
                                &                    &     &  && O3N2 &    N2  
\\
\noalign{\smallskip}
 \hline
 \noalign{\smallskip}
5.83  &    -15.01    &    1.11    &   162$\pm11$         &   263$\pm16$  &  
45$\pm3$      &  46$\pm4$       &   51$\pm5$    &     0.97       &    $-$0.76   
&     8.41  &    8.46     \\
5.10  &    -14.58    &   0.88     &   192$\pm11$ 	 &   268$\pm15$  &  
39$\pm2$      &  29$\pm2$       &   68$\pm5$    &     1.12       &    $-$0.83   
&     8.37  &    8.40    \\
4.37  &    -14.81    &   0.35     &   153$\pm9$ 	  &  279$\pm14$  &  
61$\pm4$      &  ---                  &   ---    &     0.84       &    $-$0.66  
 &     8.45  &    8.54      \\
3.65  &    -15.07    &   1.31     &   183$\pm9$ 	  &   260$\pm13$  &  
59$\pm3$      &  39$\pm3$      &  78$\pm6$    &    0.90        &    $-$0.64    &
    8.43  &    8.56    \\
2.91  &   -14.95   &    0.80    &    150$\pm8$	          &   270$\pm13$  &  
57$\pm3$      &   57$\pm4$     &   90$\pm6$   &     0.85       &    $-$0.67    &
    8.45  &     8.53   \\
2.20  &   -15.22   &    0.68    &    141$\pm8$ 	         &    272$\pm21$  &  
65$\pm3$      &   ---	           &   ---    &    0.77       &    $-$0.62    & 
   8.48  &     8.57     \\
1.45  &   -15.17   &    0.78    &    135$\pm6$ 	         &    270$\pm11$  &  
87$\pm4$      &  135$\pm7$   &    65$\pm4$   &    ---      &    ---    &     ---
&     --- \\
0.73  &   -15.25   &    1.05    &    169$\pm8$ 	         &    265$\pm10$  & 
101$\pm4$     &  142$\pm8$  &    90$\pm3$   &     ---     &    ---   &     --- 
&     ---   \\
0.38  &   -15.38   &    1.47    &    204$\pm10$	&    256$\pm10$  &  114$\pm5$   
&  136$\pm5$  &    114$\pm8$ &     ---     &    ---   &     --- &     ---  \\
0.0  &   -15.39   &    1.97   &       313$\pm12$ 	 &    247$\pm8$    & 
153$\pm6$    &  143$\pm9$  &    145$\pm11$ &     ---         &       ---        
&     ----   &      ---\\
0.38  &   -15.39   &    1.71   &     262$\pm10$ 	&    252$\pm10$   & 
154$\pm6$   &  124$\pm8$  &    132$\pm10$ &     ---         &    ---           
&       ---  &      --- \\
0.73  &   -15.21   &    1.03   &     207$\pm8$ 	         &    265$\pm9$     & 
174$\pm6$   &  113$\pm9$  &    148$\pm7$ &     ---         &    ---            &
     ---  &      --- \\
1.45  &   -15.35   &    1.68   &     352$\pm14$ 	 &   253$\pm10$   & 
152$\pm6$   &   109$\pm5$  &   152$\pm7$  &     ---        &    ---            &
     ---    &      ---  \\
2.20  &   -15.02   &    0.47   &     192$\pm7$ 	         &    276$\pm11$   & 
106$\pm5$   &   117$\pm6$  &    90$\pm7$   &     0.69      &    $-$0.41    &    
 8.50 &      8.74  \\
2.91  &   -15.11   &    0.63   &     169$\pm6$ 	         &    273$\pm11$   & 
67$\pm9$     &    ---               &   ---     &    0.83       &   $-$0.61    &
     8.46  &      8.58     \\
3.65  &   -15.27   &    0.86   &     209$\pm10$ 	&    268$\pm12$   & 
66$\pm3$     &  105$\pm6$   &    45$\pm3$   &    0.92       &   $-$0.60    &    
 8.43  &      8.58  \\
4.37  &   -15.27   &    1.44   &     137$\pm7$           &    257$\pm15$   & 
47$\pm4$     &   69$\pm4$     &   55$\pm3$   &     0.87      &    $-$0.73    &  
  8.45   &      8.48   \\
5.10  &   -15.03   &    0.53   &     124$\pm6$ 	         &    275$\pm16$   & 
55$\pm5$     &   79$\pm5$     &   16$\pm1$   &     0.79      &    $-$0.69    &  
  8.47   &      8.51   \\
\noalign{\smallskip}		 	    	     	        	  	
    		     		        	    	     
\hline
\label{tablen}
\end{tabular} 
\end{table*}

\begin{figure}
  \centering
  \includegraphics[width=\columnwidth]{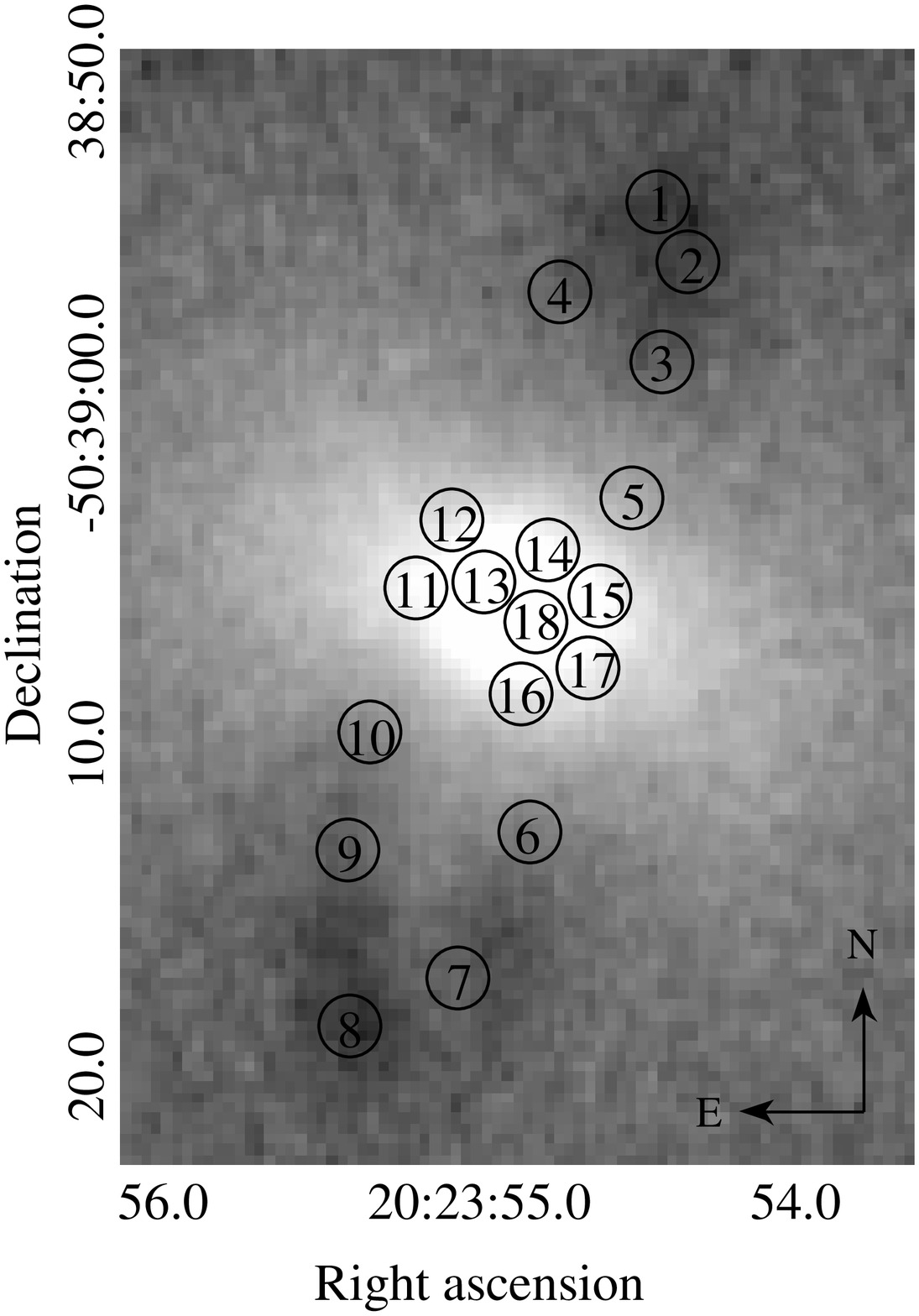}
\caption{(B-R) Color map: the blue color of the ring (dark) compared to the red
color of the galaxy (lighter). Circles mark the positions used for apperture photometry. Measured
magnitudes are presented in Table~\ref{mag}.}
  \label{foto} 
\end{figure}

\subsection{Apperture Photometry} \label{color-2}

\begin{figure}
  \centering
  \includegraphics[width=\columnwidth]{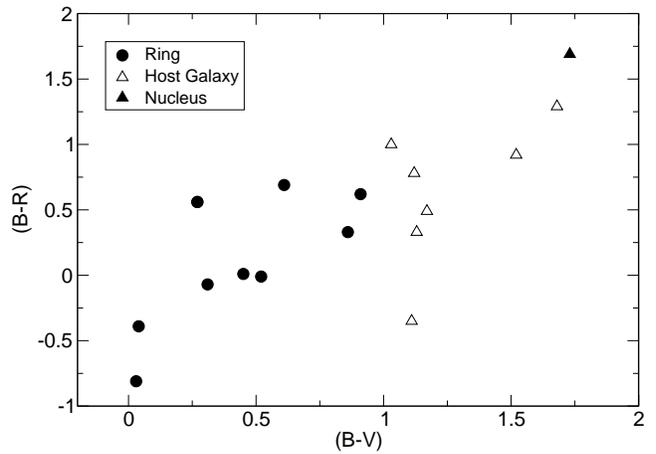}
\caption{(B-V)$\times$(B-R) diagram. Filled circles mark the ring appertures,
empty triangles correspond to the host galaxy and the filled triangle is the
nuclear region.}
  \label{color_color} 
\end{figure}

We carried out circular apperture photometry at selected positions of the
system, covering the ring, the host galaxy and its nucleus. Based in the
diagrams displayed in \mbox{Figure \ref{foto}}, in Table \ref{mag} we present
the measured magnitudes in B, (B-V), (B-R), (V-R) and (V-I) for the labeled
regions. The ring apertures have a mean (B-V) value of 0.43 mag, while the host
galaxy is  1.25 mag. The low (B-V) values on the ring, indicate that it is a
structure very different from the central galaxy: the ring is younger and of
distinct origin. In the (B-V) $\times$ (B-R) diagram, shown in Figure
\ref{color_color}, the ring, host galaxy and nucleus are very well separated.
Again, the host galaxy tends to be redder and the ring bluer. These values are
consistent with those found in other ring galaxies, as described in the case of
HRG2302 \citep{1999A&A...351..860M}. A ring bluer than the host galaxy is
expected in PRGs, because their rings are the result of recent interactions, and
they are made of material that comes from donnor galaxies, which are probably
spiral, and in this case, the material comes from its outer, less bound, parts.
These colors suggest a contribution of old stellar population in the host
galaxy, and the contribution of young stars in the ring may be due to localized
star formation (see also \citealt{1999A&A...351..860M}). 

\textbf{Colormaps:} The (B-R) color map is shown in Figure \ref{foto}. In
this greyscale map, darker regions  represent bluer colors, while lighter
regions represent redder colors. Clearly the ring is bluer than the host galaxy.
This is also seen in other colormaps, like (B-I) and (B-V).

\begin{table}
  \begin{center}
    \caption {Apperture photometry data. Apperture numbers correspond to the
positions marked in Figure~\ref{foto}.}
     \label{mag}
    \begin{tabular}{|c|c|c|c|c|c|c|}
      \hline
      \textbf{Region} & \textbf{Label} & \textbf{B} & \textbf{(B-V)} &
\textbf{(B-R)} & \textbf{(V-R)} & \textbf{(V-I)} \\
      \hline
      \hline
	
	&1	&	18.65 &0.86 & 0.33 & -0.53 & 0.14\\
	&2	& 	17.26 &0.61 & 0.69 & 0.08 & 0.03 \\
	& 3	& 	17.10 &0.45 & 0.01 & -0.44 & 0.42\\
	&4	& 	18.58 &0.31 & -0.07 & -0.38 & 0.44 \\	
Ring    &5	& 	18.42 &0.52 & -0.01 & -0.40 & 0.40 \\
	& 6	&	18.70 &0.27 & 0.56 & 0.29 & 0.0 \\
	&7	& 	18.69 &0.03 & -0.81 & -0.84 & 0.06\\	
	&8	& 	18.55 &0.04 & -0.39 & -0.43 & 0.19\\
	&9	& 	18.70 &0.27 & 0.56 & 0.29 & 0.07 \\
	&10	& 	18.12 &0.91 & 0.62 & -0.29 & 0.72 \\

\hline
	&11 & 17.14 &1.03 & 1.00 & 0.11 & 0.88\\
	&12 & 17.15 &1.12 & 0.78 & -0.34 & 0.93 \\
	&13 & 17.23 &1.52 & 0.92 & -0.60 & 0.74 \\
   Host &14 & 17.17 &1.13 & 0.33 & -0.80 & 0.8\\
Galaxy	&15 & 19.62 &1.17 & 0.49 & -0.68 & 0.0 \\
	&16 & 17.65 &1.11 & -0.35 & -1.46 & 0.34\\
	&17 & 17.69 &1.68 & 1.29 & -0.39 & 0.72 \\
\hline
Nucleus &18 & 19.70 &1.73 & 1.69 & -0.04 & 0.05 \\
\hline
    \end{tabular}
  \end{center}
\end{table}

\section{Conclusion} \label{con}

This work presents a study of \mbox{AM\,2020-504}, a galaxy with a well defined
polar ring surrounding an elliptical host galaxy (\citealt{1987IAUS..127..413W},
\citealt{1993A&A...267...21A} and \citealt{2002A&A...391..103I}). The ring was
probably formed by accretion of material from a donnor galaxy during an
interaction event. In the field around the galaxy, we did not find any nearby
object that might have given material for the formation of the ring, but there
is a group of nearby galaxies with similar radial velocities.

We estimated a redshift of \textit{z}= 0.01683, corresponding to a heliocentric
radial velocity of 5045$\pm$23 km/s, confirming the values found by
\cite{1987IAUS..127..413W} and \cite{1993A&A...267...21A}. The rotation curve of
the ring is symmetrical and well behaved. The last two points each side of the
rotation curve suggest that the northern and southern portions of the ring have
a difference in rotation velocity of about 60 km/s, but this difference is under
the error bars. To a certain degree, asymmetries could be explained if the ring
was warped.

We found the (B-R) color index averaged 0.35 and 1.73 for the ring and core of
the host galaxy respectively. Thus the ring is bluer than the host galaxy (bulge
+ nucleus), and that is what we expectif the ring is the result of a recent
interaction.

The B-band brightness profile along the minor axis of the
galaxy is asymmetric due to the ring. The NW peak is higher and corresponds to
the bright spots seen in the images. This morphological feature, as the general
S--shaped appearence of the ring are in good agreement with the warped model of
the polar ring done by \cite{1993A&A...267...21A}. The light profile along the
host galaxy major axis also looks asymmetric on both sides close to the center.
This seems to be due to the presence of dust where the ring passes in front of
the galaxy, an indication that the near side of the ring is to the NE of the
galaxy.

This system is a harbours an AGN as indicated by some diagnostic diagrams.  
Using two empirical methods based on the emission-lines easily observable, 
we found: (i) oxygen abundances  for the H\,II regions located at the ring  in
the range 12+log(O/H)=8.3-8.8 dex with an average value of $8.53\pm0.11$ dex and
(ii) the presence of an oxygen gradient across the ring of about $-0.035$
dex/kpc. We also found that  \mbox{AM\,2020-504}  follows the
metallicity-luminosity relation of typical spiral galaxies. These results
support the accretion scenario for this object and rule out cold accretion.

\section{Acknowledgements}

This work was partially supported by Universidade do Vale do Para\'{i}ba -
UNIVAP and the Minist\'erio da Ci\^{e}ncia, Tecnologia e Inova\c{c}\~{a}o
(MCTI), Laborat\'{o}rio Nacional de Astrof\'{i}sica. P. Freitas-Lemes thanks
FAPESP for the scholarship granted under process 2010/17136-4. O.L.Dors is
greateful to the FAPESP for support under grant 2009/14787-7.
  We thank the anonymous referee for helping us make this manuscript a better paper.

\bibliography{publicacao}
\bibliographystyle{mn2e}

\end{document}